# Observing the Carbon-Climate System

David Schimel, Jet Propulsion Lab, California Institute of Technology, Pasadena, CA, 91109

And,

Piers Sellers, NASA GSFC, Berrien Moore III, University of Oklahoma, Abhishek Chatterjee, NASA GSFC, David Baker, Colorado State University, Joe Berry, Carnegie Institute for Science, Kevin Bowman, NASA JPL, Phillipe Ciais LSCE, France, David Crisp, NASA JPL, Sean Crowell, University of Oklahoma, Scott Denning, Colorado State University, Riley Duren, NASA JPL, Pierre Friedlingstein, University of Exeter, UK, Michelle Gierach, NASA JPL, Kevin Gurney, Arizona State University, Kathy Hibbard, NASA HQ, Richard A. Houghton, Woods Hole Research Ctr., Deborah Huntzinger, Northern Arizona University, George Hurtt, University of Maryland, Ken Jucks, NASA HQ, Randy Kawa, NASA GSFC, Randy Koster, NASA GSFC, Charles Koven, Lawrence Berkeley National Lab, Yiqi Luo, University of Oklahoma, Jeff Masek, NASA GSFC, Galen McKinley, University of Wisconsin-Madison, Charles Miller, NASA JPL, John B. Miller, NOAA ESRL, Paul Moorcroft, Harvard University, Ray Nassar, Environment Canada, Chris O'Dell, Colorado State University, Leslie Ott, NASA GSFC, Steven Pawson, NASA GSFC, Michael Puma, Columbia University, Tristan Quaife, University of Reading UK, Haris Riris, NASA GSFC, Natasha Romanou, NASA GISS, Cecile Rousseaux, NASA GSFC, Andrew Schuh, Colorado State University, Elena Shevliakova, NOAA GFDL, Jim Tucker, NASA GSFC, Ying Ping Wang, CSIRO Australia, Christopher Williams, Clark University, Xiangming Xiao, University of Oklahoma, Tatsuya Yokota, NIES Japan

**Abstract**

Increases in atmospheric CO2 and CH4 result from a combination of forcing from anthropogenic emissions and Earth System feedbacks that reduce or amplify the effects of those emissions on atmospheric concentrations. Despite decades of research carbon-climate feedbacks remain poorly quantified. The impact of these uncertainties on future climate are of increasing concern, especially in the wake of recent climate negotiations. Emissions, long concentrated in the developed world, are now shifting to developing countries, where the emissions inventories have larger uncertainties. The fraction of anthropogenic CO2 remaining in the atmosphere has remained remarkably constant over the last 50 years. Will this change in the future as the climate evolves? Concentrations of CH4, the 2nd most important greenhouse gas, which had apparently stabilized, have recently resumed their increase, but the exact cause for this is unknown. While greenhouse gases affect the global atmosphere, their sources and sinks are remarkably heterogeneous in time and space, and traditional in situ observing systems do not provide the coverage and resolution to attribute the changes to these greenhouse gases to specific sources or sinks. In the past few years, space-based technologies have shown promise for monitoring carbon stocks and fluxes. Advanced versions of these capabilities could transform our understanding and provide the data needed to quantify carbon-climate feedbacks. A new observing system that allows resolving global high resolution fluxes will capture variations on time and space scales that allow the attribution of these fluxes to underlying mechanisms.

**Capsule Summary**

Carbon cycle feedbacks modify the effects of anthropogenic $CO_2$ on the climate, but these feedbacks are poorly understood. We propose satellite observations of carbon-climate feedbacks aimed at improving their prediction.

**Introduction**

What controls the rate of increase of the greenhouse gases carbon dioxide ($CO_2$) and methane ($CH_4$) in the atmosphere? The atmospheric concentrations of greenhouse gases (GHGs), principally $CO_2$ and $CH_4$ have increased substantially over the last century, primarily because of fossil fuel use, land use change and other anthropogenic activities. The current global annual mean atmospheric concentration of $CO_2$ exceeds 400 parts per million (ppm) and is growing at a rate of ~2 ppm/yr (± 0.1 ppm/yr). Similarly, $CH_4$ emissions have accelerated since 2007 and now exceed 1800 parts per billion (ppb), roughly a 2.5 times increase over pre-industrial levels. As a result, changes in atmospheric radiative forcing arising from greenhouse gas emissions will likely be the most important driver of climate change in the 21$^{st}$ century.

Natural processes play a significant role in controlling the atmospheric $CO_2$ and $CH_4$ abundances (Figure 1). In particular, the land biosphere and the ocean are currently absorbing over half of the $CO_2$ emitted by human activities, while chemical reactions in the atmosphere provide the primary loss mechanism for $CH_4$. Understanding the relationship between climate forcing from anthropogenic emissions, climate-carbon cycle feedbacks, and the resulting atmospheric $CO_2$ and $CH_4$ *concentrations* in a changing climate has been recognized as an important goal by the IPCC (5$^{th}$ Assessment Report, 2013). To do this, the behavior of anthropogenic carbon sources, magnitude and patterns of natural land and ocean sinks and several key processes controlling them must be quantified (Schimel et al., 1995; US Carbon Science Plan., 2011, Andres et al. 2012, Mckinley et al 2015).

While the rates of increase of GHGs are well-documented by observations, and the inputs to the atmosphere from anthropogenic emissions are relatively well-known, rates of increase of $CO_2$ and $CH_4$ are not explained by the trend in anthropogenic emissions alone. $CO_2$ concentrations would be much higher if it were not for large compensating uptakes by the terrestrial biosphere and oceans, which have offset more than 50% of anthropogenic $CO_2$ emissions to date. Models suggest the fraction of fossil emissions retained in the atmosphere will change, but these predictions vary substantially, such that carbon cycle feedbacks are one of the largest uncertainties in climate projections (Bodman et al 2013). Significant changes in the airborne fraction (Raupach et al. 2014) could drastically affect the future trajectory of $CO_2$ increases, with associated impacts on climate. The processes controlling these effects are not adequately understood nor incorporated in Earth System Models and so our predictive ability is sharply limited.

A number of scientific questions regarding human perturbations to the carbon cycle have come into focus in the past few years. Resolving these questions will reduce uncertainty and ultimately lead to improved predictions of future climate. How much fossil carbon ($CO_2$ and $CH_4$) is emitted each year, and how are these emissions distributed in time and space? $CO_2$ emissions from fossil fuel use and cement manufacturing, and $CH_4$ emissions from agriculture are now estimated on the scale of individual countries primarily from statistical and economic data (Marland et al 2009). The natural processes controlling atmospheric $CO_2$ and $CH_4$ are even less well quantified on regional to continental scales. Why do carbon cycle feedbacks on land and in the ocean absorb about half of anthropogenic $CO_2$, and will these feedbacks change in the future as the climate evolves? Because of their importance, these natural fluxes and their geographic distribution must be better quantified from observations. This is especially true in suspected "centers of action" such as the tropics, the Southern Ocean and central Asia, where few $CO_2$ measurements are collected. How is terrestrial absorption of $CO_2$ distributed geographically and what mechanisms cause this uptake? In particular, is the current terrestrial carbon uptake due to recovery from historic land use (a forcing) or does it result from changing climate and increasing $CO_2$ (a feedback). Projecting these two mechanisms into the future results in very different outcomes, but we cannot today confidently assign fluxes quantitatively to either cause.

Why is $CH_4$ increasing? Earth System feedbacks play a role in the changing atmospheric concentration of $CH_4$. The $CH_4$ growth rate slowed to near zero from 1992 to 2006, and then began to increase again (Dlugokencky et al 2011). Changes to $CH_4$ fluxes may have been due to changes in the water and carbon cycles in the tropics (Gloor et al 2013). Although evidence suggests recent impact of tropical fluxes, changes are also possible in the high latitude wetlands, which are currently poorly monitored. Even more recent increases have been attributed to forcing from the energy sector (Caulton et al, 2014) but our current observations cannot discriminate unambiguously between natural and fossil drivers of $CH_4$.

Models of carbon cycle feedbacks for both $CO_2$ and $CH_4$ disagree spectacularly (Friedlingstein et al 2006, Melton et al. 2013, Randerson et al. 2015) and current observing systems and ecosystem experiments have failed to resolve these differences (Schimel et al 2015a). The current *in situ* system of highly precise and accurate atmospheric $CO_2$ and $CH_4$ measurements constrains global budgets and provides the foundation on which all other carbon observing systems are based. The *in situ* network does not allow fluxes to be estimated at fine enough scales to adequately constrain process-based carbon cycle models and allow diagnosis and/or attribution of the carbon fluxes with any confidence, especially in the tropics or high latitudes. The existing surface greenhouse gas observing networks are not configured to target the extended, complex and most dynamic regions in the carbon cycle. Global networks of carbon flux observations provide detailed process information across a range of biomes at local scales, but don't provide sufficient coverage to scale up to regional or global estimates.

Both anthropogenic and natural carbon cycle processes have high spatial and temporal variability at local scales, but the global climate forcing depends on the spatially- and temporally-integrated impact of these processes, as the carbon cycle responds to coherent changes in regional climates, such as El Nino droughts, and other factors. Interannual climate variation affects $CO_2$ and $CH_4$ fluxes at regional and sub-seasonal scales. These offer natural experiments that can be used to understand the sensitivities of fluxes to climate. Because of high variability at small spatial and temporal scales, the "scaling-up" of local observations to global scales is difficult, and requires a multi-scale measurement and modeling framework. Space-based observations hold the promise of resolving the crucial intermediate scales, observing processes over regions large enough to link carbon cycle processes to the global climate system.

Testing models globally, not at individual study sites but across the world's major regions, is a necessary step in falsifying inadequate models, elucidating the physical, chemical and biological processes acting, and reducing uncertainty in projections into the future (Sitch et al. 2015). Large differences between model predictions, and our inability to adequately benchmark and test these models against the current suite of available carbon cycle measurements, severely limits their usefulness for detecting important changes and improving our understanding of the predictability of the Earth's carbon cycle. ***A new observing system that quantifies fluxes globally, particularly in regions with high fluxes or reservoirs of carbon, and, captures the scales of variation in time and space that allow attribution to underlying mechanisms is urgently required***. The new observing system should resolve fluxes at the same scales as interannual climate variation to be able to estimate consistent sensitivities.

Confounding these scientific challenges, uncertainty in fossil emissions is actually growing. Fossil emissions, once concentrated in a relatively few countries with mature reporting systems, are now widely distributed around the world, including high emissions from several rapidly evolving economies. Technologies for production, distribution and use of fossil fuels are also changing, adding another layer of complexity. Uncertainty is not only increasing in absolute magnitude as emissions increase, but also the relative uncertainty is growing due to lags in reporting. An increase in fossil fuel uncertainty leads to greater uncertainty in the global carbon budget and causes increased uncertainty about the magnitude of present-day feedbacks as well, since they are typically quantified as residuals after the better known-fluxes are accounted for (e.g., the Global Carbon Project - Le Quéré et al. 2015).

There is a clear need to better understand and predict future carbon-climate feedbacks, so that science can more confidently inform climate policy, including adaptation planning and future mitigation strategies. Rising concentrations of atmospheric $CO_2$ and $CH_4$, and their interplay with the land and ocean carbon cycles and the climate system, remains the primary source of uncertainty. Recognizing the importance of this issue, a community workshop was held at the University of Oklahoma on March 16-18, 2015 to identify the outstanding questions and translate them into observing requirements, in preparation for the next Decadal Survey of the Earth Sciences and Applications from Space, by the National Research Council. This paper reports the primary scientific findings from that group.

**Why a space-based approach?**

One way to reduce the uncertainty in carbon cycle-climate feedbacks is to collect high-resolution observations of $CO_2$ and $CH_4$ concentrationsrom space-based measurement platforms (Figure 2). These data can provide information at the time and space scales required for testing hypotheses about the relevant continental and ocean basin-scale processes. Emerging measurement systems, including the Japanese Greenhouse gases Observing SATellite (GOSAT), the NASA Orbiting Carbon Observatory-2 (OCO-2), the ESA Sentinel 5p, and other new sources of data provide a first step towards reducing these uncertainties. These spacecraft collect thousands to hundreds of thousands of measurements of the column averaged dry air mole fractions of $CO_2$ ($X_{CO2}$) and $CH_4$ ($X_{CH4}$) over the sunlit hemisphere each day. However, these pioneering missions do not provide the spatiotemporal coverage to answer the key carbon-climate feedback questions at process-relevant scales or address the distribution and quantification of anthropogenic sources at urban scales. Nevertheless, they do demonstrate that a well-planned future system integrating space-based and *in situ* observations and measurements could provide the accuracy, spatial resolution, and coverage needed to address these issues (Ciais et al., 2014; CEOS Report, 2014).

While the key observables are $CO_2$ and $CH_4$ *concentrations*, the goal is to provide "top-down" or estimates of the surface-atmosphere *fluxes* of $CO_2$ and $CH_4$ targeting 100 km (~1°x1°) and monthly scales, although these finest scales may not be achievable in the cloudiest regions. The fluxes are estimated by combining satellite data and in situ measurements with atmospheric flux inversion models. The 100 km target can resolve many key ecosystem and land use processes. For major urban areas, and for estimation of anthropogenic emissions, the flux determinations need to be at spatial scales on the order of 10 km, with concomitant shorter time scales. Measurement of proxy trace gases, such as carbon monoxide, CO, would help significantly on the issue of anthropogenic source attribution (fossil fuel and biomass combustion, in particular). A sustained time series is required to quantify interannual variability and trends and to estimate the climate sensitivity of fluxes. For example, the ENSO cycle currently drives substantial variation in terrestrial and marine carbon cycles (Wang et al 2015, Anderegg et al. 2015). To capture the variability of El Nino forcing, and separate the various climatic influences on the carbon cycle, such as temperature and precipitation, at least two and ideally more El Nino events should be captured, requiring a decade or so of uninterrupted observations.

Pieter Tans famously observed that "the carbon cycle is one" but carbon measurements have remained persistently stovepiped, with programs aimed at land, ocean or atmospheric observations and having limited interaction. Atmospheric observations bring unity to the study of the carbon cycle, as they observe the imprint of the, land, ocean and anthroposphere on the atmosphere, and space-based observations provide the most comprehensive and spatially resolved view of the atmosphere. Addressing the critical uncertainties in the global carbon cycle requires a coordinated effort with synergistic and calibration/validation studies for all these domains. We note that there are certain key Earth System properties, for example carbon in soils or the deep ocean, that cannot be observed from space with any known technology, but are critical and synergistic with the space-based program, beyond basic calibration/validation requirements (Schimel et al., 2015a). A few of these are highlighted in the following section.

**Observing Carbon Fluxes Globally**

On average, over the past decade or so,, the anthropogenic combustion of fossil fuels is releasing about 10 billion metric tons of carbon per year (or Gigatons of carbon as $CO_2$ per year or GtC/yr) while land use change is adding an additional 0.9 GtC/yr to the atmosphere (Table 1). Terrestrial photosynthesis takes up around 120 GtC/yr and also removes about one quarter of the anthropogenic $CO_2$ emissions. The 120 GtC/yr in photosynthesis is however nearly balanced by respiration, leaving a small net terrestrial sink of 2.5 GtC/yr or so.  Similarly, the oceans emit nearly 80 GtC/yr (IPCC AR5, 2013) and reabsorb this amount plus another quarter of the anthropogenic emissions. The land and ocean fluxes vary widely with climate from year to year, so that in any given year, the amount of fossil emissions remaining in the atmosphere can also vary.  This variability is the clearest signal of the climate sensitivity of the carbon cycle and is a key focus for research.  On average, though, the land and ocean fluxes result in slightly less than 50% of anthropogenic $CO_2$ emissions remaining in the atmosphere, leading to a contemporary average global $CO_2$ growth rate of ~2 ppm/yr (± 0.1 ppm/yr).

Global emissions of $CO_2$ from fossil fuel combustion (and cement manufacture) currently have an uncertainty of ~5%. The uncertainty in fossil-fuel emission from many developed countries is typically less than 5%, while the uncertainty in fossil-fuel emissions from recently developed countries (such as China, India, Brazil and Russia) is 10-20%; whereas, in much of the developing world the larger uncertainties are offset by the magnitude of the emissions being smaller (Marland et al., 2009; Guan et al., 2012; Andres et al., 2012). As the fraction of global emissions originating from rapidly expanding and newly developed economies as well as from lesser-developed nations has grown, so has the uncertainty in global fossil fuel emissions. As mentioned earlier, a major reason for reducing this uncertainty is that it propagates through the inferred fluxes in atmospheric inversion calculations and increases the resultant uncertainty from feedbacks. At sub-national scales, the consequences of such uncertainty on inflow into the domain of interest can have a critical impact on the estimates of local biospheric emissions and uptake.

Along with $CO_2$, quantifying the emissions of $CH_4$ is critical for projecting and mitigating changes to climate. After a period of rapid increase in the 1980's, atmospheric $CH_4$ concentrations stabilized for almost a decade between the mid-1990s and the mid-2000s (Kai et al., 2011), and have begun to increase rapidly since 2007 (Rigby et al., 2008; Dlugokencky et al. 2009; Nisbet et al., 2014). Wetland emissions, mainly from freshwater wetlands, represent the largest and most uncertain natural source of $CH_4$, with published estimates ranging from 140 to 280 million tons $CH_4$ per year (Bloom et al., 2010; Kirschke et al., 2013). Emissions from other, smaller sources, such as coastal marshes, termites, permafrost, and freshwaters are documented by only a handful of studies. Anthropogenic sources including wet (rice paddy) agriculture, biomass burning, and $CH_4$ leaks from the coal, oil and gas extraction and transport (Caulton et al. 2014), landfills, waste-water processing, are also poorly known. At present, they are mainly estimated from statistical data on energy use and emission factors, both of which vary greatly. But again our current observational capabilities cannot discriminate unambiguously between natural and fossil drivers of $CH_4$.

The atmosphere has a central and integrating role in the global carbon cycle as the primary medium for carbon exchange between the larger land and ocean carbon reservoirs, and this role, as noted earlier, implies that atmospheric measurements of $X_{CO2}$ and $X_{CH4}$ can constrain $CO_2$ and $CH_4$ flux calculations. Via advanced modeling techniques such as inverse modeling and data assimilation frameworks (Enting 2002), these atmospheric observations are then used to provide carbon flux estimates over large areas (the "top-down" approach). Substantial uncertainty is still contributed to inversion models from their representation of atmospheric transport, especially convection and boundary layer diffusion, and improvements in these physics are also needed. On the other hand, complex physical and biological processes that underlie the carbon cycle can only be understood using the "bottom-up" approach. Top-down flux estimates can be directly compared with bottom-up estimates of the fluxes generated from carbon cycle models forced by local environmental and remotely sensed data (see Figure 3) to define the attribution of sinks and sources, and thereby resolve model ambiguities. Because the bottom-up modeling captures process information, reconciliation of the two approaches offers the potential to advance understanding of the underlying processes (and their parameterizations in ESMs) and enable fundamental advances and increases in predictive capacity (Schimel et al., 2015b) for the carbon-climate system, supporting both basic geophysical understanding and policy-relevant

12 | P a g e

applications.

Recognizing the merit of such activities, recently community-wide inter-comparison efforts have been undertaken as part of the REgional Carbon Cycle Assessment and Processes (RECCAP; Canadell et al., 2012-2014) project. Results from these studies highlight that reconciling the carbon flux estimates not only minimizes the uncertainty within each approach but also increases confidence in the results from these approaches. Consequently, having robust techniques for incorporating atmospheric observations into carbon cycle inverse modeling/data assimilation frameworks is essential for connecting top-down flux estimates to improvements in the structure and parameterizations of process-based models.

**Land and ocean processes in the carbon cycle**

Estimating regional fluxes is crucial to quantify feedbacks, but additional measurements are also relevant to partition net fluxes into the contributing component fluxes, such as photosynthesis and respiration, or ocean uptake versus outgassing. The terrestrial biosphere plays a major role in moderating atmospheric $CO_2$ increase, but stocks of carbon on land depend on the balance between photosynthetic uptake, respiration and fire, all processes that respond quite differently to climate. Observations of carbon uptake, losses, and associated changes in stocks in the terrestrial biosphere are all key for assessing the causes of uptake and release of land carbon in the coupled carbon-climate system.

Gross uptake of carbon by photosynthesis ("gross primary production" or GPP) is the source of carbon to the land biosphere and is sensitive to climate change (Anav et al. 2015). Carbon loss through respiration from plants and microbes, together with uptake via GPP defines the carbon balance of an undisturbed ecosystem but responds to climate differently from GPP. Several missions (GOME, GOSAT, OCO-2 and 3, and the new European FLEX) measure a new quantity, Solar-Induced Fluorescence (SIF) which is a direct by-product of photosynthesis, and provides a unique new constraint on carbon uptake (Frankenberg et al 2011). Respiration is perhaps the least well understood of the terrestrial biological fluxes (Ryan and Law 2005). Improved direct estimates of GPP from satellite measurements, coupled with measurements of Net Ecosystem Exchange (NEE) or Production (NEP) inferred from satellite-derived $X_{CO2}$, will provide a key constraint on model formulations of respiration. Separating GPP from respiration may be possible by combining a SIF constraint on GPP with inverse estimates of NEE, allowing stronger attribution of flux anomalies to the underlying stresses and corresponding physiological responses.

Biomass measurements also provide a long-term integral constraint on land carbon fluxes. Repeated measurements of biomass provide a decadal-scale constraint on land-atmosphere models. Contemporary and planned missions provide radar and LIDAR measurements of carbon stocks in the current era. Repeat measurements on a five- to ten-year cycle, and with coverage at all latitudes, could monitor disturbance and land-use impacts (Schimel et al., 2015a, Hall et al 2011), provide information about the response of terrestrial biomass to climate change and increasing atmospheric $CO_2$, and help disentangle the impacts of climate (changing temperatures, precipitation patterns, etc.) from $CO_2$ fertilization [Friedlingstein 2006].

While the land biosphere is currently a net sink for $CO_2$, it is a net source of $CH_4$. Methane fluxes arise from fossil fuel production and distribution, agriculture, wetlands, and fire. The contributions of each of these are poorly understood and consequently ecosystem process models show very little agreement in their estimates of $CH_4$ fluxes (Melton et al., 2013). While process models will continue to rely on bottom-up parameterizations based on inventories and fire emission data, remote sensing products such as soil moisture, inundation by surface water, gravity anomalies and land use, provide an opportunity to refine models of $CH_4$ emissions. A step-change in understanding could be obtained by collecting new, systematic atmospheric measurements of $CH_4$ concentrations.

Ocean carbon exchange is likewise a critical feedback to atmospheric $CO_2$, responding to both changing concentrations and climate. Carbon fluxes between the ocean and atmosphere are regulated primarily by the gradient between atmospheric and oceanic $pCO_2$. Ocean $pCO_2$ is set primarily by circulation and two main oceanic processes: the solubility and the biological pumps. The solubility pump is the largest regulator, and is directly affected by changes in physical circulation, surface winds and stratification of the water column. Increasing temperatures and dissolved inorganic carbon (DIC) concentrations in oceanic surface waters are expected to decrease the efficiency of the solubility pump thereby slowing down the uptake of atmospheric $CO_2$.

Satellite measurements of physical ocean variables such as ocean topography and temperature are essential to estimating the time-evolving air-sea $CO_2$ exchange (Brix et al 2015), but chemical and biological measurements made *in situ* are also required. Since the 1970s, such data have come primarily from research ships, but now are increasingly available from autonomous instruments on commercial ships, moorings, and gliders. Currently, major efforts are underway to deploy autonomous drifting floats with multiple biogeochemical sensors as an extension of the ARGO program that has used floats to make physical measurements since 2003. ARGO floats with $pCO_2$ sensors are currently under development and will provide an additional constraint on the net ocean flux (Fiedler et al., 2013). Detailed biogeochemical measurements from research ships from top to bottom in the ocean are critical for validation of autonomous approaches, as well as for tracking long-term absorption of carbon by the ocean. Measurements of surface ocean $pCO_2$ from both research and commercial ships will also continue to be very important for validation of monthly to annual ocean fluxes as inferred from both bottom-up and top-down techniques (Roedenbeck et al. 2015).

Ocean biology and biogeochemistry also affect surface ocean $pCO_2$ and drive carbon to the deep ocean for long-term storage. Phytoplankton absorbs carbon dioxide during photosynthesis, which converts inorganic carbon into organic carbon, a small portion of which eventually sinks to the deep oceans. These processes can be indexed using ocean color measurements (Siegel et al 2014). Current and planned measurements (e.g. the NASA PACE mission) of ocean color, a proxy for phytoplankton concentration, provide a constraint on phytoplankton productivity as well as other key ocean processes of the biological pump. Coastal-zone carbon fluxes are poorly quantified and mechanistic understanding is limited, but current estimates suggest that they may represent 25% of the total ocean $CO_2$ absorption (Benway et al. 2016, Landschützer et al., 2014, Laruelle et al., 2014). Future climate and land-use changes could significantly impact these fluxes.

**Space-based measurement techniques**

Measurements of $X_{CO2}$ and $X_{CH4}$ can be retrieved from high-resolution spectroscopic observations of reflected sunlight in near infrared $CO_2$ and $CH_4$ bands by dividing by the total column of air that is similarly obtained from $O_2$ measurements. As mentioned earlier, these concentrations can then be used to estimate fluxes via atmospheric inverse models. Measurements using active LIDAR remote sensing are also possible and are currently being explored using aircraft. Space-based remote sensing observations of these gases are challenging because these gases are so long-lived that even strong sources or sinks produce changes of small magnitude relative to the background concentrations. Consequently, for these measurements to be useful for deriving information about surface fluxes, they need to be made with extremely high precision and high accuracy or low biases (Rayner and O'Brien, 2001). JAXA's GOSAT mission for $X_{CO2}$ and $X_{CH4}$ and NASA's OCO-2 mission for $X_{CO2}$ have pioneered this capability, but a much denser grid of concentration observations is needed to retrieve estimates of surface carbon fluxes at the target spatial and temporal resolution (100 km or ~1°x1°, and monthly). An important ancillary space measurement is carbon monoxide (CO), which helps to separate emissions from various combustion sources and other processes such as biospheric respiration. *In situ* studies of CO/$CO_2$ ratios from different types of burning (liquid, gas or solid fossil fuels or biomass) show promise in disentangling emissions from different sectors. When the atmospheric $CO_2$ information is combined with the information from CO as an additional trace gas species significant improvement in $CO_2$ surface flux attribution can be obtained (Wang et al., 2009).

There are several space-borne measurement approaches for observing atmospheric $X_{CO2}$, $X_{CH4}$ and $X_{CO}$ concentrations that are likely to be available during the next decade. The relative strengths and weaknesses of these approaches have not yet been thoroughly examined. The current missions provide broad-scale constraints on fluxes, but increasing the density of observations even more could transform our understanding of the carbon cycle, by allowing fluxes to be resolved at scales at which underlying biogeochemical processes become evident. A carefully selected combination of $CO_2$, $CH_4$ and CO measurements from both Low Earth Orbit (LEO) and Geostationary (GEO) missions may be needed since they observe complementary time and space scales. LEO can give global sampling with a single mission, albeit with coverage and repeat interval dictated by instrument swath width. Active LEO systems offer potential advantages in terms of reduced bias and low-illumination coverage, especially over high-latitudes.

GEO platforms provide complementary information over large but fixed regions, allowing fluxes to be determined on shorter time scales and over finer spatial scales and providing detail not available from the LEO vantage point. A constellation of GEO platforms could provide persistent observations at higher frequency over all land between ~55ºS and 55ºN latitudes with benefits for observing megacities and would acquire far more data in cloudy environments. Geostationary-like observations of $CO_2$, $CH_4$ and CO at higher latitudes (>55°) could be made by observing from a highly elliptical orbit (HEO) (Nassar et al., 2014). It is not yet clear what combination of techniques and platform configurations would yield the best science results for a given level of investment. Hence, it is essential to explore the measurement trade space rigorously prior to any detailed discussions about mission specification, including the needed contributions from surface networks and airborne measurements.

The oceanic, terrestrial and atmospheric considerations stated above assume the continuation of a wide range of ancillary and supporting measurements for calibration/validation and retrieval algorithm development activities. For example, *in situ* vertical profiles of trace gases are required to calibrate ground-based remote sensing sites to the WMO $CO_2$, $CH_4$ and CO scales used in the surface networks, which enables calibration of the space-based measurements to the same absolute scale. Cross-calibration between international missions will ensure high quality, complementary and consistent data records of atmospheric carbon species. Refinements to existing retrieval algorithms need to focus on improvement of $CO_2$, $CH_4$ and CO spectroscopy and treatment of clouds, aerosols and other sources of bias. Along with the above, in situ observational data streams are needed to provide a long-term stable context to interpret the space-based measurements of greenhouse gases. These data improve the accuracy and range of validity of remote sensing retrieval algorithms and facilitate the attribution of the estimated fluxes to specific natural and anthropogenic processes.

Vigorous model advancement, benchmarking, and inter-comparison activities need to be continued and expanded in parallel with improved observations to reduce uncertainty about feedbacks in the future. Improvements in both process-based and atmospheric transport models are needed. Coordination with ongoing and planned air quality observations may help capture some of the high-resolution transport features: in fact, integrating air quality with carbon cycle science can advance both areas. Improved integration and coordination of interdisciplinary observation and modeling tools are critical for answering the overarching science questions laid out at the beginning.

**Conclusions**

There is no reason to expect that relationship between emissions, climate and the resulting atmospheric concentrations of $CO_2$ and $CH_4$ will remain relatively constant over the next few decades, as has largely been the case over the last century. While the centennial behavior of the carbon system has been surprisingly consistent, we know of no guarantee from theory or the paleorecord that this will continue as temperatures rise. To take one example, it is quite possible that the land biosphere will change from being a strong net sink of $CO_2$ to being a much weaker one, or even a source in the future. Currently, there are very few observational constraints on the dynamics of these global feedbacks and their evolution with the changing climate.

The potential for significant feedback effects on the future trajectory of increasing $CO_2$ is large, changing the relationship between emissions and climate, and thus changing the calculus of mitigation efforts. If negative feedbacks weaken or positive feedbacks strengthen, then climate impacts could be more severe than anticipated, approaching or exceeding dangerous interference levels. If positive feedbacks are lower than projected, then mitigation efforts might lose economic efficiency. This science area must be studied and feedback effects included in future models to make them more accurate. Testing and improving the land-surface and ocean parameterizations in ESMs that calculate the surface-atmosphere fluxes of energy, water, and carbon, is essential to improve their predictive capability.

Next-generation missions can transform our understanding of carbon-climate feedbacks, but the relative benefits of different measurement strategies need to be explored thoroughly. It is clear that several satellite platforms will be required to achieve the target of estimating monthly 1° x 1° fluxes globally from top-down inversions. The optimal combination of sensor types (active, passive systems), and platform orbits (Low Earth Orbit/LEO; Geostationary Orbit/GEO; Highly Elliptical Orbit/HEO) has not been quantified. Achieving the necessary carbon cycle science from a coordinated suite of observations will require international collaboration and continued investment in calibration and validation, including the surface observing system. Investments now in simulations (i.e., OSSEs) and analyses can determine the most cost-effective observing system that meet the science needs. These OSSEs will require time and resources, but this undertaking is necessary to build and fly the right instruments, make these challenging measurements and achieve the desired science goals.

Each of these developments and research efforts provide strong constraints on one or more aspects of the carbon cycle, and weaker constraints on others, and all have different costs and risks. The time has come to clearly articulate the most critical scientific uncertainties, identify the observations critical to reducing those uncertainties, and use these insights to prioritize the optimum systems for the next decade of carbon science. The vision for a carbon-climate observing system must be shared internationally through organizations such as Committee for Earth Observations from Space, the Global Carbon Observing System, the Group on Earth Observations and the World Meteorological Organization, as partnerships between the space-faring nations will likely be needed for its realization. Developing a next-generation observing capability for the carbon cycle requires close interaction between observationalists, modelers and technologists, but will lead to both fundamental discovery and improved forecasting and management of the global carbon and climate systems.

**Acknowledgements**. The authors gratefully acknowledge the support of Lee Anne Sallee with logistics and the University of Oklahoma for support and hospitality. The project was partially supported by a NASA award to OU. The research carried out at the Jet Propulsion Laboratory, California Institute of Technology, was under a contract with the National Aeronautics and Space Administration. Copyright 2016 California Institute of Technology. This document is provided by the contributing author(s) as a means to ensure timely dissemination of scholarly and technical work on a noncommercial basis. Copyright and all rights therein are maintained by the author(s) or by other copyright owners. It is understood that all persons copying this information will adhere to the terms and constraints invoked by each author's copyright. This work may not be reposted without explicit permission of the copyright owner.

**Figure captions**

*Figure 1. Schematic showing how anthropogenic emissions (dark gray box) are translated into concentrations by feedbacks from ecosystems and the ocean, currently reducing concentrations, and hence climate forcing, considerably below what they would be otherwise . The uptake of a fraction of emissions by land and ocean feedbacks modulates the climate impact of each unit of emission, since the climate system responds to changes in concentration. Several of the processes believed to be most important and likely to change--human forcing, tropical and high latitude carbon storage and changing ocean physics and biogeochemistry--are emphasized.*

*Figure 2. Two seasons of the carbon cycle. Global maps of the column-average $CO_2$ dry air mole fraction ($X_{CO2}$) in northern hemisphere Fall and Spring 2015, produced from OCO-2 observations. The color bar extends from 390 ppm (blue) to 405 ppm (red).*

*Figure 3. How are fluxes linked to underlying controls? Estimates of surface carbon fluxes can be inferred via two main approaches - "bottom-up" (based on measurements in the land or ocean components, and process-based models) and "top-down" (based on measurements in the atmosphere). Being able to reconcile the estimates from these two approaches and combine them in an integrated framework provides consistent results and explanation across scales and allows for validation, attribution and prediction. Shown here are the desired estimation scales for obtaining the surface carbon fluxes – one focused on regional fluxes and another focused on urban anthropogenic emissions at finer spatial and temporal scales.*

*Table 1. The global carbon budget, shown for one example year (2014). In this budget, the net land flux is shown, and includes emissions from deforestation, which are more than balanced by uptake. Future feedbacks could affect the relative magnitude of the emission and uptake fluxes, leading to an altered retention of fossil $CO_2$ in the atmosphere. From http://www.globalcarbonproject.org/carbonbudget/15/hl-compact.htm.*